\documentclass[a4paper]{article}

\usepackage[english]{babel}
\usepackage[utf8x]{inputenc}
\usepackage[T1]{fontenc}

\usepackage[a4paper,top=3cm,bottom=2cm,left=3cm,right=3cm,marginparwidth=2cm]{geometry}

\usepackage{amsmath}
\usepackage{graphicx}
\usepackage[colorinlistoftodos]{todonotes}
\usepackage[colorlinks=true, allcolors=blue]{hyperref}
\usepackage{authblk}

\usepackage[round,sort,comma,numbers]{natbib}

\title{Frequency Domain Multiplexing for MKIDs: Comparing the Xilinx ZCU111 RFSoC with their new 2x2 RFSoC board}

\author[a,b,*]{Eoin Baldwin}
\author[a,b]{Mario De Lucia}
\author[c, b]{Colm Bracken}
\author[a, b]{Gerhard Ulbricht}
\author[a]{Oisin Creaner}
\author[a, b]{Jack Piercy}
\author[a, b]{Tom Ray}

\affil[a]{Dublin Institute for Advanced Studies, Astronomy \& Astrophysics, 31 Fitzwilliam Place, Dublin 2, Ireland}
\affil[b]{Trinity College Dublin and CRANN, School of Physics, Dublin 2, Ireland}
\affil[c]{Maynooth University, Department of Experimental Physics, Maynooth, Co Kildare, Ireland}

\begin{document}
\maketitle

\begin{abstract}
The Xilinx ZCU111 Radio Frequency System on Chip (RFSoC) is a promising solution for reading out large arrays of microwave kinetic inductance detectors (MKIDs). The board boasts eight on-chip 12-bit / 4.096 GSPS analogue-to-digital converters (ADCs) and eight 14-bit / 6.554 GSPS digital-to-analogue converters (DACs), as well as field programmable gate array (FPGA) resources of 930,000 logic cells and 4,272 digital signal processing (DSP) slices. While this is sufficient data converter bandwidth for the readout of 8,000 MKIDs, with a 2 MHz channel-spacing, and a 1 MHz sampling rate (per channel), additional FPGA resources are needed to perform the DSP needed to process this large number of MKIDs. A solution to this problem is the new Xilinx RFSoC 2x2 board. This board costs only one fifth of the ZCU111 while still providing the same logic resources as the ZCU111, albeit with only a quarter of the data converter resources. Thus, using multiple RFSoC 2x2 boards would provide a better balance between FPGA resources and data converters, allowing the full utilization of the RF bandwidth provided by the data converters, while also lowering the cost per pixel value of the readout system, from approximately €2.50 per pixel with the ZCU111, to €1 per pixel. 
\end{abstract}

\section{Introduction}
\begin{figure}
\centering
\includegraphics[width=0.5\textwidth]{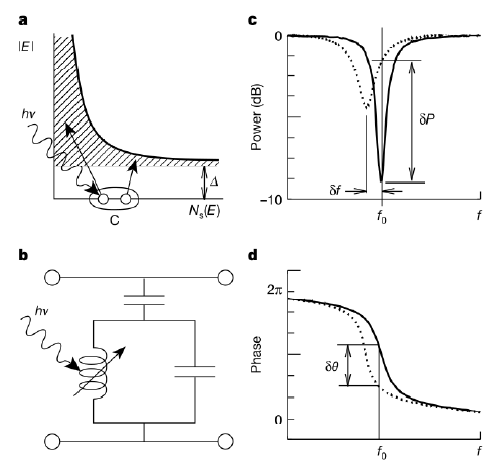}
\caption{\label{fig:detectionprinciple}MKID Detection Principle}
\end{figure}

Microwave Kinetic Inductance Detectors (MKIDs) are low temperature detectors based on superconducting LC resonators, and the kinetic inductance effect. Figure \ref{fig:detectionprinciple} displays their basic detection principle \cite{dayarticle}.  Incident photons which strike the detector's inductor break superconducting Cooper pairs, which causes an increase in quasi-particles. This increase in quasi-particle density results in a corresponding increase in inductance. This causes the resonant frequency of the resonator to shift by: 

\begin{equation}
    f_0 = \frac{1}{\sqrt{LC}}
\end{equation}

When monitoring a single frequency, this shift in resonant frequency can be seen as either a change in the amplitude or the phase of the resonator's transmission. Because this shift is proportional to the number of quasi-particles formed, which is proportional to the energy of the incident photon, the phase shift is also proportional to the energy of the incident photon. This means that the energy of incident photons can be determined by monitoring the phase of an MKID. This gives MKIDs their intrinsic energy resolution, which is one of their key benefits.

While other low temperature detectors, such as transition edge sensors (TESs) or metallic magnetocalorimeters (MMCs), also have intrinsic energy resolution, they have the disadvantage of being difficult to scale up to large arrays, needing the use of Superconducting Quantum Interference Devices (SQUIDs) to be multiplexed \cite{squids_bates}. MKIDs, however, can be designed as an array of resonators, whereby each resonator has a unique resonant frequency, suitably spaced from nearby frequencies, allowing the whole array to be read out on a single feedline by using frequency domain multiplexing (FDM). 

In terms of low temperature electronics, an array of thousands of MKIDS requires only a single HEMT amplifer on a single feedline. The trade off to this simplicity at low temperatures is the need for room temperature electronics capable of performing the complex digital signal processing (DSP) needed to separate the whole bandwidth of the array into a series of channels, with one channel for each pixel in the array.

\begin{figure}
\centering
\includegraphics[width=0.8\textwidth]{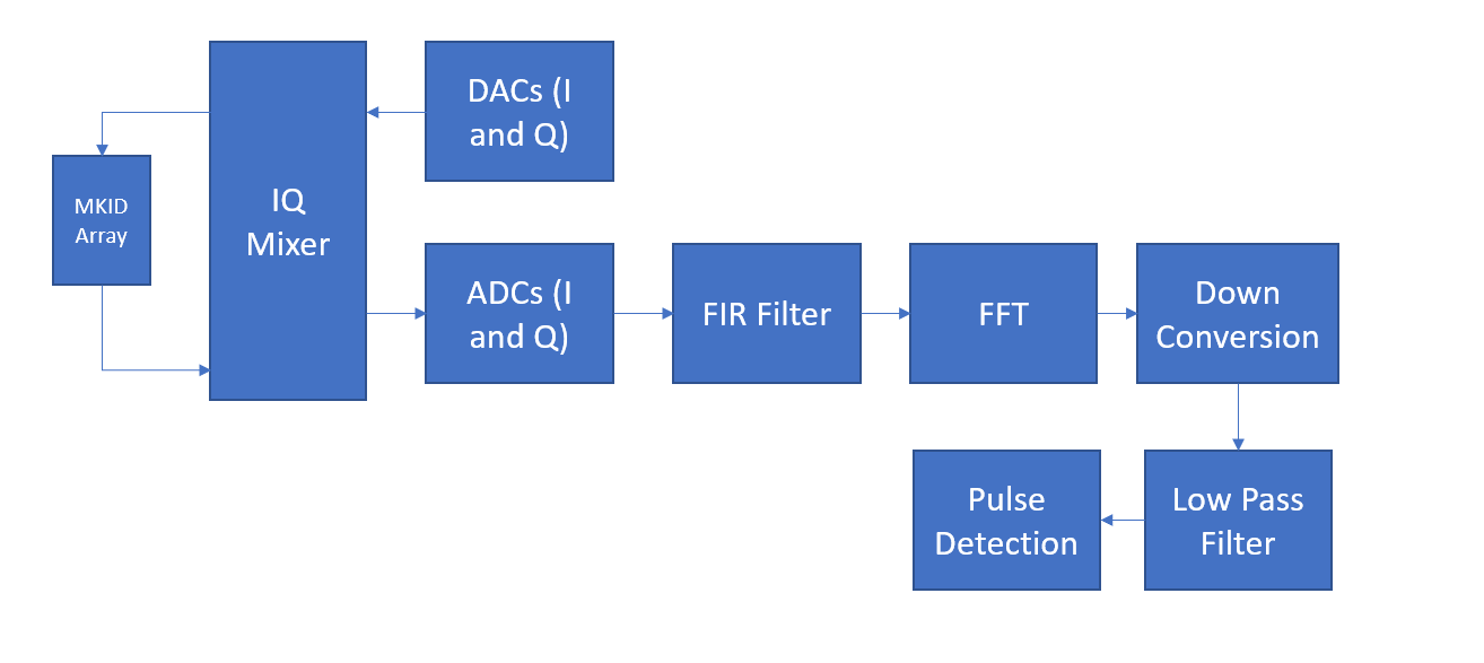}
\caption{\label{fig:readoutimplementation}Possible MKID Readout Implementation}
\end{figure}

To readout an array of MKIDs, digital-to-analogue converters (DACs) are used to generate an array of frequencies at, typically, MHz frequencies. An intermediate frequency (IF) board then upmixes these tones to GHz frequencies, such that there is a single tone for each MKID in the array, with the frequency of each tone being equal to that of the corresponding MKID. The upmixed signal passes through the MKID array, and is amplified at low temperatures by a HEMT amplifier, and at room temperature by a room temperature amplifier. After being amplified at room temperature, these signals are downmixed back to MHz frequencies by the IF board, before being finally re-digitized by analogue-to digital-converters (ADCs). An FPGA board is then used to perform the channelization DSP on this digitized data. Channelization divides the entire frequency span into a series of frequency bands, such that each frequency band contains a single pixel in the MKID array. Each of these bands is then monitored for phase pulses, signifying photon events \cite{straderthesis}. Fig \ref{fig:readoutimplementation} displays a possible implementation of an MKID readout system. 

Thus, in order to carry out this signal processing for large arrays of MKIDs, room temperature electronics consisting of DACs, ADCs, an FPGA and an IF mixer board are required. MKID readout systems which have already been implemented to read out large arrays of pixels include the ROACH 1 \cite{mchughreadout} and ROACH 2 \cite{Fruitwala_2020} boards. However, for even larger, next generation arrays of MKIDs, more powerful FPGA boards, and higher frequency data converters are required. One promising solution which is being explored in DIAS, is the Xilinx ZCU111 Radio Frequency System-on-Chip (RFSoC). \cite{rfsocdatasheet}

\section{Xilinx ZCU111 RFSoC} \label{zcu111section}

\begin{figure}
\centering
\includegraphics[width=1\textwidth]{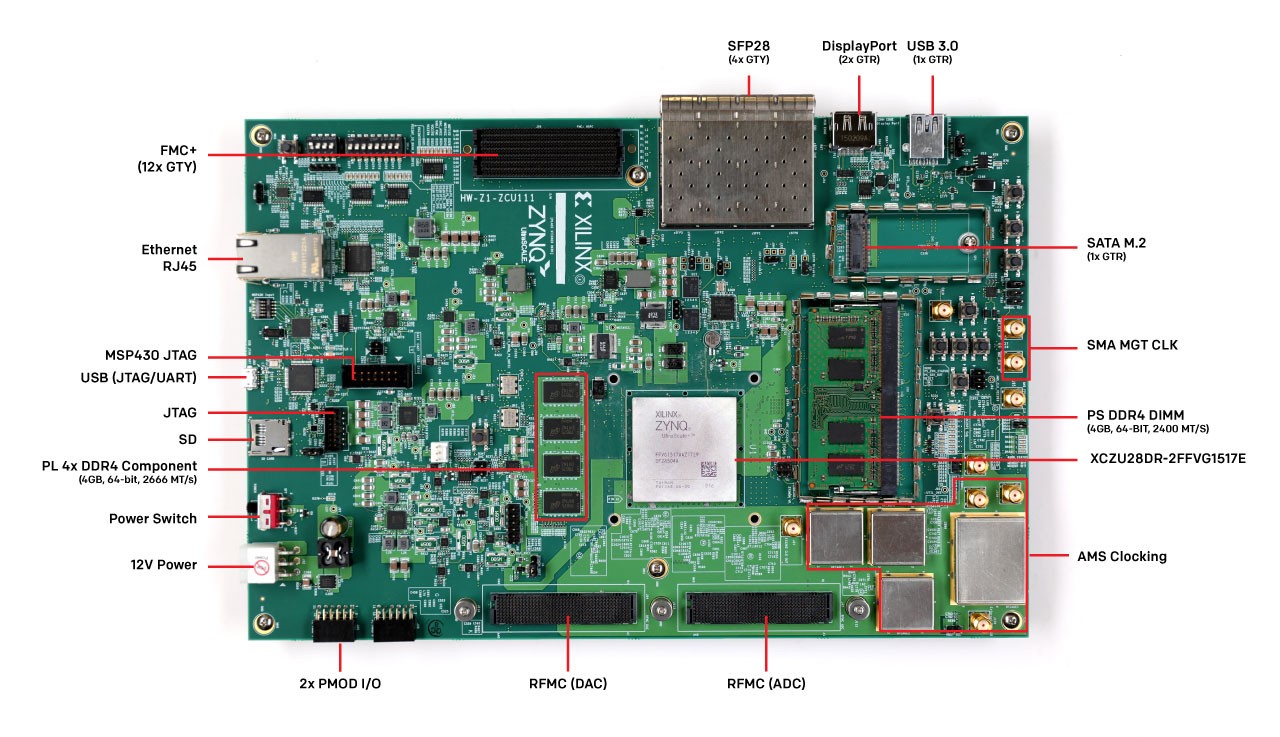}
\caption{\label{fig:zcu111}Xilinx ZCU111 RFSoC}
\end{figure}

The ZCU111 is a radio frequency system-on-chip (RFSoC) developed by Xilinx. It boasts FPGA resources of 4,272 DSP slices, and 930,000 logic cells. What is particularly appealing about this board, is that it does not require external data converters, having eight on-chip 12 bit / 4.096 GSPS ADCs, and eight on-chip  14 bit / 6.554 GSPS DACs, on the same system-on-chip (SoC) as the FPGA. Not needing external data converters greatly reduces the size of this MKID readout system. Each feedline requires two ADCs, and two DACs, in order to sample both the I and Q components of the resonators transmission. Thus, each feedline, requires a pair of ADCs and a pair of DACs. With an ADC sampling rate of 4.096 GSPS, and a pixel spacing of 1 MHz, this allows for 2,000 pixels to be read out on a single feedline, with the possibility for four feedlines. Thus, the ZCU111 has sufficient data converters resources to read out upto 8,000 MKIDs. However, while this is ample data converter resources to read out an array of 8,000 pixels, the FPGA resources are estimated to only be sufficient for processing approximately 4,000 pixels. Thus, in order to fully utilize the RF bandwidth provided by the ZCU111's data converters, additional FPGA resources are required. \cite{rfsocdatasheet}  

\section{Xilinx 2x2 RFSoC}

\begin{figure}
\centering
\includegraphics[width=0.5\textwidth]{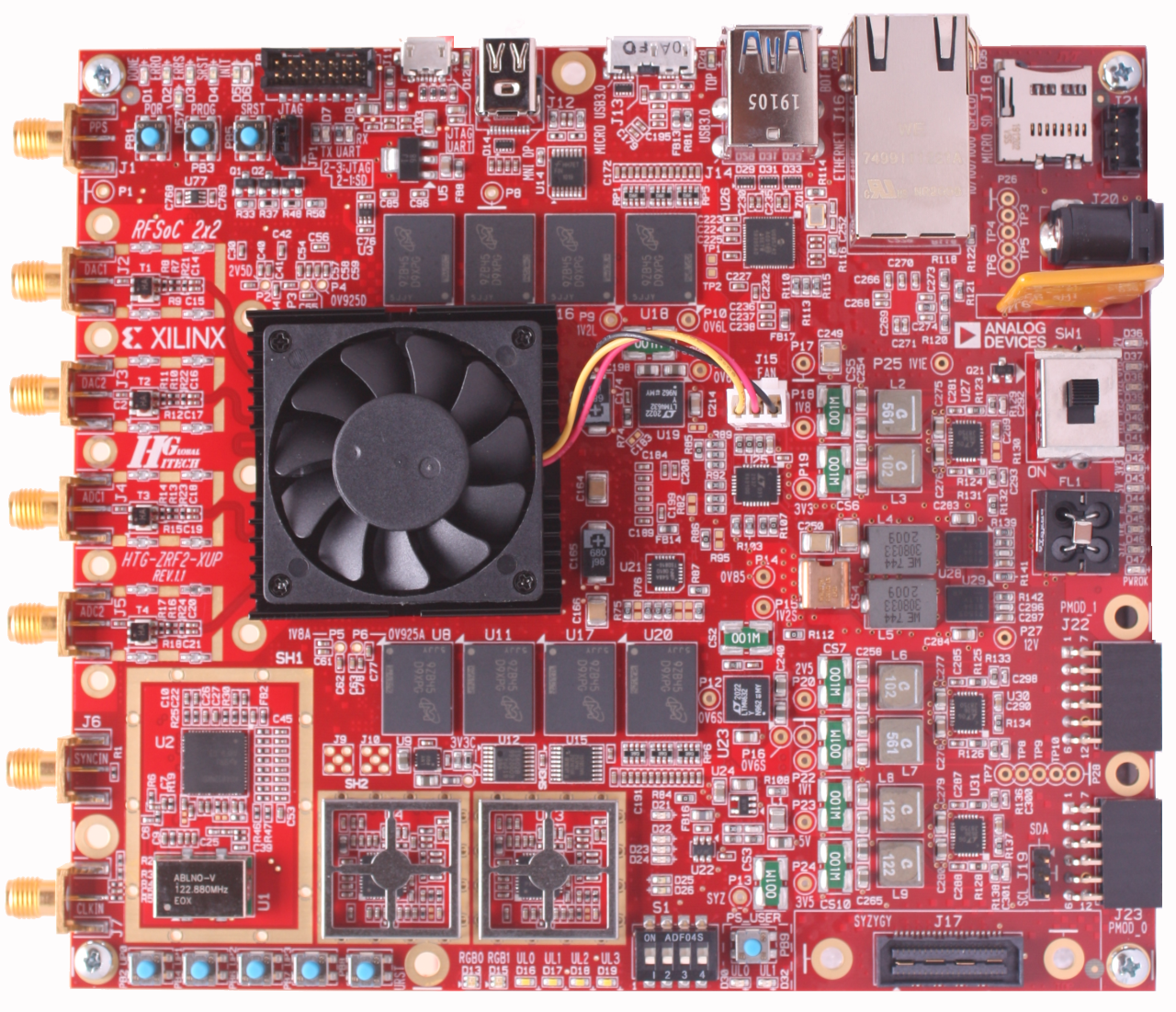}
\caption{\label{fig:2x2}Xilinx 2x2 RFSoC}
\end{figure}

The 2x2 RFSoC \cite{rfsoc2x2datasheet} is a low cost alternative to the ZCU111 released by Xilinx as part of their University Programme. While it contains the same FPGA resources as the ZCU111, it boasts only a quarter of the data converter resources, containing only two on-chip 12 bit / 4.096 GSPS ADCs, and two on-chip  14 bit / 6.554 GSPS DACs. Importantly, it is also only a fifth of the price, driving down the cost-per-pixel value of a potential MKID readout system. As it has an abundance of FPGA resources, relative to data converters resources, the full 4.096 GSPS bandwidth of the ADC/DAC pairs can be used, meaning that each board can read out 2,000 pixels.  

\section{Cost Per Pixel Values}
One important metric which is used to evaluate MKID readout systems is their cost-per-pixel value. For example, the ARCONS\cite{arcons} and DARKNESS\cite{meekertthesis} arrays used ROACH 1 and ROACH 2 boards, respectively, to read out arrays of optical MKIDs, with cost per pixel values shown in Table \ref{previousboards}.  

\begin{table}[] 
\caption{FPGA Board Comparison} \label{previousboards}
\begin{tabular}{p{3.2cm}p{1.4cm}p{1.8cm}p{1.2cm}p{1.8cm}}
\hline
 \textbf{Board} & \textbf{Total Cost (€)} & \textbf{Pixel Count} & \textbf{Cost/Pixel (€/pixel)} \\ \hline
 ROACH 1 & 3,770 & 256 & 14.73 \\ \hline
 ROACH 2 & 3,850 & 1,000 & 3.85 \\ \hline
\end{tabular}
\end{table}

As explained in Section \ref{zcu111section}, while the ZCU111 has ample data converter resources for reading out upto 8,000 pixels, the actual number of MKIDs which it is estimated to be capable of reading out is confined by the FPGA resources to ~4,000 pixels. This, is based on scaling up the pixel counts achieved by ROACH 1 and ROACH 2 relative to their FPGA resources. Thus, a single ZCU111, costing ~€10,000, has a cost per pixel of €2.50. 

Initially, it was proposed that an external FPGA board could be coupled to the ZCU111 to provide additional FPGA resources, in order to fully utilize the full bandwidth of the RF data converters, driving down the cost per pixel. However, the 2x2 RFSoC, having four times the FPGA resources relative to data converter resources, when compared to the ZCU111, allows for the entire RF bandwidth of the data converters to be used. Thus, a single 2x2 RFSoC, capable of reading out 2,000 pixels, and costing ~€2,000, will have a cost per pixel of €1. This is summarised in Table \ref{rfsocboards}.

\begin{table}[] 
\caption{RFSoC Board Comparison} \label{rfsocboards}
\begin{tabular}{p{3.2cm}p{1.4cm}p{1.8cm}p{1.2cm}p{1.8cm}}
\hline
 \textbf{Board} & \textbf{Total Cost (€)} & \textbf{Pixel Count} & \textbf{Cost/Pixel (€/pixel)} \\ \hline
 ZCU111 & 10,000 & 4,000 & 2.50 \\ \hline
 2x2 RFSoC & 2,000 & 2,000 & 1.00 \\ \hline
\end{tabular}
\end{table}

\section{Development Progress}
Work has been ongoing in developing firmware and software for the ZCU111 for reading out arrays of MKIDs. Moreover, a pair of 2x2 RFSoC's have been ordered, and development will continue on these. As both boards use the same FPGA, adapting the firmware and software from the ZCU111 to the 2x2 should be relatively straightforward. 

The firmware for the board is developed using Xilinx's Vivado package, while the accompanying software is written in Python, using the PYNQ framework. PYNQ uses Jupyterlab Python, running on the RFSoC, to create user friendly notebooks for writing to and reading from the Vivado generated firmware. 

\begin{figure}
\centering
\includegraphics[width=1\textwidth]{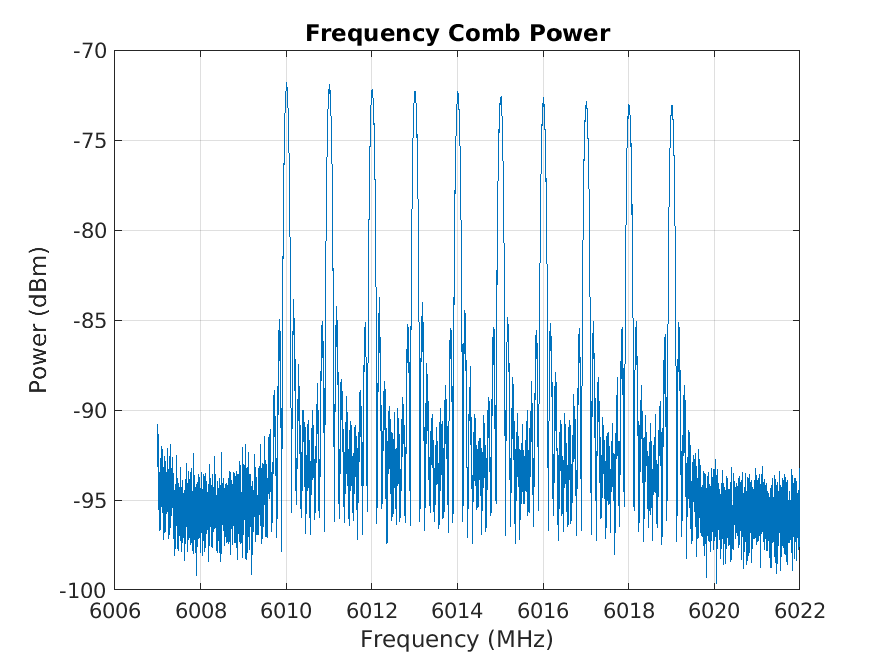}
\caption{\label{fig:freqcomb}Frequency Comb Generated at MHz Frequencies and Upmixed to 6 GHz}
\end{figure}

\begin{figure}
\centering
\includegraphics[width=1\textwidth]{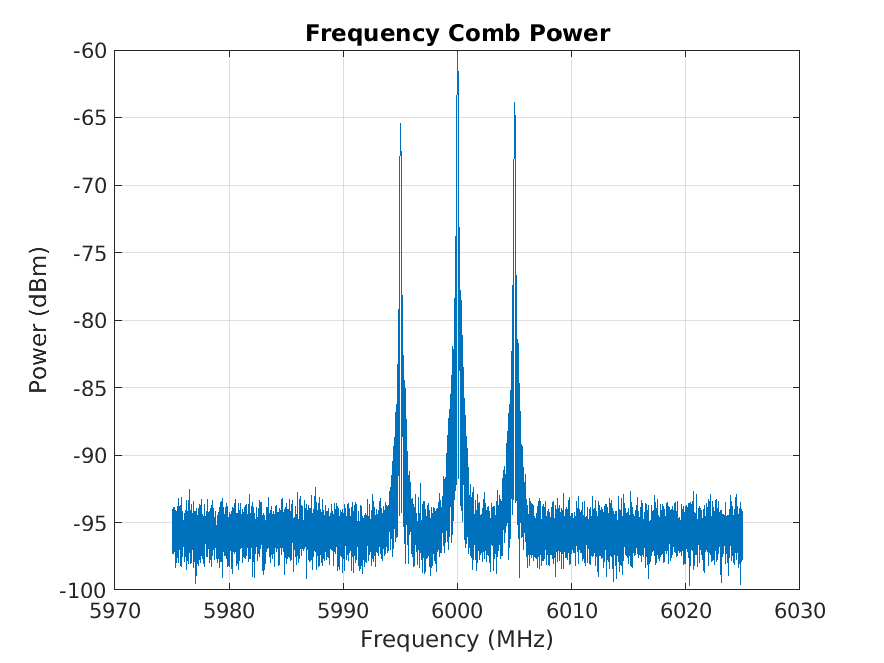}
\caption{\label{fig:singletone}Single Tone Upmixed to Either Side of 6 GHz Local Oscillator}
\end{figure}

Current firmware developed for the ZCU111, using Xilinx Vivado,  can be used to generate frequency combs using the board's DACs, which are upmixed to GHz frequencies using an intermediate frequency (IF) mixer board. Similarly, signals which are downmixed back to MHz frequencies by the same mixer board, are digitized by the boards ADCs. Programmable finite impulse response (FIR) filters and fast Fourier transforms (FFT) are then applied to these digitized signals, separating the frequency span into a series of channels. Clocked at 409.6 MHz, the signals are, typically, channelized using a $2^{10}$ tap FFT, and 25 tap bandpass low pass FIR filter. Figure \ref{fig:freqcomb} shows a frequency comb generated by the ZCU111's DACs at MHz frequencies and upmixed with the 6 GHz local oscillator by the IF mixer board, and measured using a vector network analyzer (VNA). Figure \ref{fig:singletone} shows a single 5 MHz tone, upmixed to either side of the LO, giving the upper and lower sidebands.  

\section{Conclusion}
Work has been done developing firmware and software for the Xilinx ZCU111 RFSoC, implementing code to control the board's data converters, in order to generate frequency combs, which can then be upmixed with an external mixer board, as well as implementing FFT and filtering blocks needed to perform the channelization needed to read out arrays of MKIDs. Next, it is intended to also implement this firmware on the new Xilinx 2x2 RFSoC to provide a cheaper readout alternative. 

\bibliographystyle{unsrtnat}
\bibliography{eoins_sources}

\begin{thebibliography}{9}
\providecommand{\natexlab}[1]{#1}
\providecommand{\url}[1]{\texttt{#1}}
\expandafter\ifx\csname urlstyle\endcsname\relax
  \providecommand{\doi}[1]{doi: #1}\else
  \providecommand{\doi}{doi: \begingroup \urlstyle{rm}\Url}\fi

\bibitem[Day et~al.(2003)Day, Leduc, A~Mazin, Vayonakis, and
  Zmuidzinas]{dayarticle}
Peter Day, Henry Leduc, Benjamin A~Mazin, Anastasios Vayonakis, and Jonas
  Zmuidzinas.
\newblock A broadband superconducting detector suitable for use in large
  arrays.
\newblock \emph{Nature}, 425:\penalty0 817--21, 10 2003.
\newblock \doi{10.1038/nature02037}.

\bibitem[{Mates}(2011)]{squids_bates}
J.A.B {Mates}.
\newblock \emph{{The Microwave SQUID Multiplexer}}.
\newblock PhD thesis, University of Colorado, 2011.

\bibitem[{Strader}(2016)]{straderthesis}
Matthew~James {Strader}.
\newblock \emph{{Digital Readout for Microwave Kinetic Inductance Detectors and
  Applications in High Time Resolution Astronomy}}.
\newblock PhD thesis, University of California, Santa Barbara, 2016.

\bibitem[McHugh et~al.(2012)McHugh, A~Mazin, Serfass, Meeker, O'Brien, Duan,
  Raffanti, and Werthimer]{mchughreadout}
Sean McHugh, Benjamin A~Mazin, Bruno Serfass, Seth Meeker, Kieran O'Brien, Ran
  Duan, Rick Raffanti, and Dan Werthimer.
\newblock A readout for large arrays of microwave kinetic inductance detectors.
\newblock \emph{The Review of scientific instruments}, 83:\penalty0 044702, 04
  2012.
\newblock \doi{10.1063/1.3700812}.

\bibitem[Fruitwala et~al.(2020)Fruitwala, Strader, Cancelo, Zmuda, Treptow,
  Wilcer, Stoughton, Walter, Zobrist, Collura, and et~al.]{Fruitwala_2020}
Neelay Fruitwala, Paschal Strader, Gustavo Cancelo, Ted Zmuda, Ken Treptow,
  Neal Wilcer, Chris Stoughton, Alex~B. Walter, Nicholas Zobrist, Giulia
  Collura, and et~al.
\newblock Second generation readout for large format photon counting microwave
  kinetic inductance detectors.
\newblock \emph{Review of Scientific Instruments}, 91\penalty0 (12):\penalty0
  124705, Dec 2020.
\newblock ISSN 1089-7623.
\newblock \doi{10.1063/5.0029457}.
\newblock URL \url{http://dx.doi.org/10.1063/5.0029457}.

\bibitem[Xil(2018)]{rfsocdatasheet}
\emph{ZCU111 Evaluation Board User Guide}.
\newblock Xilinx, 10 2018.

\bibitem[Hit(2021)]{rfsoc2x2datasheet}
\emph{RFSoC 2x2 User Manual}.
\newblock Hitech Global, 1 2021.

\bibitem[A.~Mazin et~al.(2013)A.~Mazin, R.~Meeker, Strader, abd Bumble,
  O'Brien, Szypryt, Marsden, C.~van Eyken, E.~Duggan, Ulbricht, Stoughton, and
  Johnson]{arcons}
B~A.~Mazin, S~R.~Meeker, Matthew Strader, B~abd Bumble, Kieran O'Brien,
  P~Szypryt, D~Marsden, J~C.~van Eyken, G~E.~Duggan, G~Ulbricht, C~Stoughton,
  and M~Johnson.
\newblock Arcons: A 2024 pixel optical through near-ir cryogenic imaging
  spectrophotometer.
\newblock \emph{Publications of the Astronomical Society of the Pacific}, 125,
  06 2013.
\newblock \doi{10.1086/674013}.

\bibitem[{Meeker}(2017)]{meekertthesis}
S.~{Meeker}.
\newblock \emph{{DARKNESS: The First Microwave Kinetic Inductance Detector
  IntegralField Spectrograph for Exoplanet Imaging}}.
\newblock PhD thesis, University of California, Santa Barbara, 2017.

\end{thebibliography}

\end{document}